\documentclass{article}
\usepackage[latin1]{inputenc}	
\usepackage{graphicx}
\begin{document}
\setlength{\textwidth}{27pc}

\title{
\footnotesize\hspace{9.5cm}UCL-IPT-01-16\\
\bigskip
\bigskip
\bigskip
\bigskip
\bigskip
\bigskip
\bigskip
\LARGE
\textbf{The Lemaître-Schwarzschild Problem Revisited}}
\normalsize
\bigskip
\bigskip
\bigskip

\author{A.FÜZFA$^{\dagger}\footnote{email: \textit{afu@math.fundp.ac.be}}$  , 
  J.-M. GERARD$^{\ddagger}$  
and D.LAMBERT$^{\ddagger}$
\\
\\
\small
$^{\dagger}$Unité de Systèmes Dynamiques,\\
\small
Facultés Universitaires N.-D. de la Paix,\\
\small
B-5000 Namur\\
\\
\small
$^{\ddagger}$Institut de Physique Théorique\\
\small
Université Catholique de Louvain\\
\small
B-1348 Louvain-la-Neuve}

\maketitle
\bigskip
\bigskip
\bigskip
\bigskip

\begin{abstract}
The Lemaître and Schwarzschild analytical solutions for a relativistic
spherical body of constant density are linked together through the use of the Weyl
quadratic invariant. The critical radius for gravitational collapse of an incompressible fluid
is shown to vary continuously from $9/8$ of the Schwarzschild radius to the
Schwarzschild radius itself while 
the internal pressures become locally anisotropic.
\end{abstract}
\pagebreak
\section{Introduction}
As early as in 1933, G. Lemaître \cite{lemaitre} emphasized that the most general stress-energy tensor 
associated with a spherical distribution of matter  is 
locally 
aniso- tropic, i.e. expressed in terms of different radial ($p_r$)
and tangential ($p_t$) pressures. This generalization which goes beyond the standard perfect fluid approximation ($p_r =p_t$) has been extensively used in the recent literature about equilibrium \cite{equilibrium} and collapse \cite{collapse}
of very compact objects. In particular, the Lemaître case of
a collapse with vanishing radial pressure ($p_r=0$ ; $p_t\neq 0$) nicely illustrates how physical (naked) singularities might naturally arise in the context of classical general relativity.\\
\\
In this note, we will restrict ourselves to the study of equilibrium conditions
for an incompressible, spherical object sustained
by different
radial and tangential pressures (with $p_t\ge p_r$). This is what we call the Lemaître-Schwarzschild problem.
First, we briefly remind the reader of two analytic solutions ($p_t=p_r$
and $p_r=0$, respectively) to outline the generic behaviour expected from a dominant tangential pressure ($p_t\ge p_r$).
Then, we introduce the useful quadratic Weyl invariant to link these two extreme solutions and to determine the real nature of the so-called Schwarzschild singularity. This relevance of the Weyl tensor
in the study of gravitational collapse is therefore
complementary to previous works on conformally flat anisotropic spheres 
\cite{herrera}.\\
Finally, we illustrate numerically how the boundary of a compact sphere can
vary continuously from $9/8$ of the Schwarzschild radius to the Schwarzschild radius itself, without breaking the equilibrium conditions. We confirm therefore
that the gravitational redshift of the radiation emitted at the surface of a supradense inhomogeneous star could, in principle, be unbounded.\\

\section{Equilibrium of locally anisotropic spheres}
The interior metric for a static spherically symmetric distribution of matter
can be written as\footnote{We will conventionally take the god-given units where $c=1\cdot$}
\begin{equation}
\label{met1}
ds^{2}=e^{2\nu(r)}dt^{2}-e^{2\lambda(r)}dr^{2}-r^{2}d\Omega^{2}
\end{equation}
where $r$ is the radial Schwarzschild coordinate and $d\Omega$, the solid angle element ($d\Omega^{2}=d\theta^{2}+\sin^{2}\theta d\varphi^{2}$).\\
As first shown by Lemaître \cite{lemaitre}, the most general spherical distribution of matter bounded by gravitation is \underline{locally} anisotropic.
This remarkable feature can be directly deduced from the null divergence of
the stress-energy tensor
\begin{equation}
\label{div}
T_{\alpha\;\;\;|\beta}^{\beta}=0
\end{equation}
which describes an ``anisotropically sustained'' body\footnote{We prefer this expression to the frequently used
''\textit{pressure anisotropy}'' which could bring the confusion 
that the pressure would depend on angular variables while, in fact, 
the radial pressure is just locally different from its tangential counterpart.}

\begin{equation}
\label{lem}
T_{\alpha}^{\beta}=diag(\rho(r),-p_r (r),-p_t (r),-p_t (r)),
\end{equation}
with 
\begin{equation}
\label{tov}
p_t- p_r=\frac{r}{2}\left[p_{r}'+(\rho+p_r)\nu'\right],
\end{equation}
the so-called Tolman-Oppenheimer-Volkoff equation. In fact, this relation (\ref{tov})
directly derives from the radial component of Eq.(\ref{div}) but can also 
be obtained from the well-known Einstein equations\footnote{Two of the authors (D.L. and A.F.) have shown how to include the effect of the cosmological
constant as a relique of an underlying bosonic string theory (cf.\cite{cordes}). But,
we will not consider that general case here as we have proved that it was not consistent with data of precessing pulsars binary systems.} relating the metric (\ref{met1}) to the stress-energy tensor (\ref{lem}) (see \cite{gravitation} for an explicit writing of these three equations).
\\
Lemaître only treated the special case of vanishing radial pressure $p_r$
in his favourite (elliptic) geometry. The full integration of the three
independent Einstein differential equations actually requires state equations
for the density and pressures, as well as boundary conditions. For recall,
the famous exterior metric
\begin{equation}
\label{schext}
ds^{2}=\left(1-\frac{r_S}{r}\right)dt^{2}-\left(1-\frac{r_S}{r}\right)^{-1}dr^{2}-r^{2}d\Omega^{2}
\end{equation}
with 
\begin{equation}
\label{rs}
r_S=2M,
\end{equation}
the Schwarzschild horizon associated with the sphere of mass $M$ and radius
$r_1$, results from the $\rho(r>r_1)=p(r\ge r_1)=0$ state equations and 
$g_{\mu\nu} (r\rightarrow\infty) =\eta_{\mu\nu}$ boundary conditions ($G=c=1$).\\
Here, the hypothesis of uniform energy density $\rho(r\le r_1)=\rho_{0}$
together with a regular metric at the origin of coordinates allow us to integrate one among the the three differential equations to obtain
\begin{equation}
\label{lambda}
e^{-2\lambda}=1-\frac{r^{2}}{\mathcal{R}^{2}}
\end{equation}
with
\begin{equation}
\label{bigr}
\mathcal{R}^{2}=\frac{r_{1}^{3}}{r_S}\cdot
\end{equation}
However, we are still left with two differential Einstein equations involving
the function  $\nu(r)$, namely
Eq.(\ref{tov}) and
\begin{equation}
\label{nup}
\nu'=\frac{1}{2}\left(1+3\frac{p_r}{\rho_0}\right)\lambda'\cdot
\end{equation}
Consequently, one additional equation of state for the radial and (or) tangential pressures is needed to completely determine the interior metric defined by
Eq.(\ref{met1}) and to analyse the equilibrium conditions of positive and finite pressures
\begin{equation}
\label{equi}
0\le p_{r,t}<\infty\cdot
\end{equation}

\subsection{The Schwarzschild perfect fluid (1916) : $p_r=p_t$}
If we now assume locally isotropic pressures, Eqs.(\ref{tov}) and (\ref{nup})
can easily be integrated to give
\begin{equation}
\label{nusch}
e^{\nu}=\displaystyle{\frac{3}{2}\left(1-\frac{r_{S}}{r_1}\right)^{\frac{1}{2}}-\frac{1}{2}\left(1-\frac{r^{2}}{\mathcal{R}^{2}}\right)^{\frac{1}{2}}}
\end{equation}
and
\begin{equation}
\label{psch}
p_r(r)=\rho_{0}\displaystyle{\frac{\left(1-\frac{r^2}{\mathcal{R}^2}\right)^{1/2}-\left(1-\frac{r_S}{r_1}\right)^{1/2}}{3\left(1-\frac{r_S}{r_1}\right)^{1/2}
-\left(1-\frac{r^2}{\mathcal{R}^2}\right)^{1/2}}}\cdot
\end{equation}
The integration constants are indeed fixed by imposing to recover the exterior
Schwarzschild solution (see Eq.(\ref{schext})) at the boundary $r=r_1\cdot$
The equilibrium condition (see Eq.(\ref{equi})) for the spherical distribution
of a perfect fluid is obviously violated at $r=0$, once the radius $r_1$ reaches a critical value located slightly above the Schwarzschild radius $r_S$ :
\begin{equation}
\label{had}
r_{1}^{MIN}=\frac{9}{8}r_S\cdot
\end{equation}
Below this minimal value, the central region of the sphere begins to collapse
such that the redshift of the radiation emitted at the surface
of the sphere cannot exceed the critical value of $2$.

\subsection{The Lemaître vaults (1933) : $p_r=0$ ; $p_t\neq 0$}

In a similar way, if the material sphere is only sustained by its tangential pressures, Eqs.(\ref{tov}) and (\ref{nup}) imply
\begin{equation}
\label{enu1}
e^{2\nu(r)}=\left(1-\displaystyle{\frac{r_{S}}{r_{1}}}\right)^{\frac{3}{2}}\;\left(1-\frac{r^{2}}{\mathcal{R}^{2}}\right)^{-\frac{1}{2}},
\end{equation}
and
\begin{equation}
\label{pt1}
p_t(r)=\frac{\rho_{0}r^{2}}{4\mathcal{R}^{2}}\left(1-\frac{r^{2}}{\mathcal{R}^{2}}\right)^{-1}\cdot
\end{equation}
So, as already stated by Lemaître in the early thirties, one can build a sphere only supported by transverse tensions, at the manner of vaults in
architecture.\\
At this point, it is historically fair to recall his analytical result \cite{lemaitre}
\begin{equation}
\label{lemaitre1}
p_t(r)=\frac{\rho_{0}}{4}\tan^{2}(\chi)
\end{equation}
where $\chi$ is a coordinate used to locate a point on the $S^{3}$ 
space. The trivial change of variables $r\rightarrow \mathcal{R}\sin(\chi)$
allows us to convert Eq.(\ref{pt1}) into Eq.(\ref{lemaitre1}).\\
These equations tell us that, here, the tangential pressure never vanishes except at the origin and is singular at the boundary when $r_1$ tends to $r_S$. Consequently, the critical radius
for equilibrium is now the Schwarzschild radius itself:
\begin{equation}
\label{eq17}
r^{MIN}_1=r_S\cdot
\end{equation}
\subsection{From Schwarzschild to Lemaître : $p_t\ge p_r$}
If the tangential pressure is everywhere smaller than the radial one, then the boundary condition $p_r(r_1)=0$ implies $p_t(r_1)<0$ and thus nothing could prevent the sphere of radius $r_1$ from collapsing.\\
On the other hand, if $p_t\ge p_r$, we expect a smooth transition from 
$\frac{9}{8}r_S$ to $r_S$ for the minimal radius, when the radial pressure tends
to zero. Our expectation is indeed supported by the corresponding smooth
evolution of the maximal redshift factor
\begin{equation}
\label{zmax}
z_{MAX}=e^{-\nu(r_{1}^{MIN})}-1
\end{equation}
of the radiation emitted at the critical surface. As already stated,
this critical factor
is equal to $2$ in the locally isotropic fluid approximation (Schwarzschild).
But Eqs.(\ref{enu1}) and (\ref{eq17}) tell us now that this measurable quantity
diverges
when the pressure $p_r$ goes to zero (Lemaître). Would this mean that we only
face a naked singularity in the extreme case of Lemaître's vaults?
To answer this question, one usually considers the behaviour of the quadratic invariant associated with the Riemann tensor, the so-called Kretschmann scalar.
\\
For the Schwarzschild inner solution, the Kretschmann scalar is given by
\begin{equation}
\label{kret1}
\left(R_{\mu\nu\rho\sigma}R^{\mu\nu\rho\sigma}\right)_{S}=
\displaystyle{\frac{3}{\mathcal{R}^{4}}\left\{5+6\frac{p_r}{\rho_0}+9\frac{p_{r}^{2}}{\rho_{0}^{2}}\right\}}\cdot
\end{equation}
On the other hand, for the Lemaître solution, we obtain
\begin{equation}
\label{kret2}
\left(R_{\mu\nu\rho\sigma}R^{\mu\nu\rho\sigma}\right)_{L}=
\displaystyle{\frac{3}{\mathcal{R}^{4}}\left\{5+4\frac{p_t}{\rho_0}+12\frac{p_{t}^{2}}{\rho_{0}^{2}}\right\}}\cdot
\end{equation}
So, 
in these two extreme models, a pole in the Kretschmann scalar simply reveals the
seat of infinite pressure when the boundary of the material sphere reaches the minimal radius allowed for equilibrium. It is therefore tempting
to conjecture that this remarkable feature holds true in all the $p_t>p_r$
intermediate models. The use of the quadratic invariant associated with the
Weyl component of the Riemann tensor turns out to be a rather powerful tool
to prove that a quasi-static collapse is indeed triggered when the radius
of the sphere lies somewhere between $\frac{9}{8}r_S$ and $r_S\cdot$
\section{The Weyl quadratic invariant}
With our choice of the metric signature $(+,-,-,-)$, 
the Weyl tensor, defined as the non-Ricci irreducible component of the Riemann tensor, is given by the following formula \cite{hawking}:
\begin{eqnarray}
\label{weyl}
W_{\mu\nu\rho\sigma}=R_{\mu\nu\rho\sigma}+\frac{1}{d-2}\left(g_{\mu\rho}R_{\nu\sigma}-g_{\mu\sigma}R_{\nu\rho}+g_{\nu\sigma}R_{\mu\rho}-g_{\nu\rho}R_{\mu\sigma}\right)\nonumber\\
-\frac{R}{(d-1)(d-2)}\left(g_{\mu\rho}g_{\nu\sigma}-g_{\mu\sigma}
g_{\nu\rho}\right).
\end{eqnarray}
This tensor is traceless and non-trivial if the dimension $d$ of space-time
is at least equal to four. It is responsible for tidal effects in
general relativity but vanishes, for example, in all the Friedmann-Lemaître-Robertson-Walker homogeneous Universes.\\
The Weyl quadratic invariant associated with the static metric given in Eq.(\ref{met1}) reads
\begin{equation}
\label{weyl1}
\mathcal{W}=W_{\mu\nu\rho\sigma}W^{\mu\nu\rho\sigma}
=\frac{4}{3}\left\{\kappa (p_r-p_t-\rho)+3\left(\frac{1-e^{-2\lambda}}{r^{2}}\right)\right\}^2
\end{equation}
with $\kappa=8\pi G\cdot$ If we deal again with an incompressible fluid of
uniform density $\rho_0$, Eq.(\ref{lambda}) implies the remarkably simple
relation
\begin{equation}
\label{weyl2}
\mathcal{W}=\frac{12}{\mathcal{R}^{4}}\left(\frac{p_r-p_t}{\rho_0}\right)^2
\end{equation}
expressing the fact that the pressure difference $p_r -p_t$ induces then local tidal
effects in the anisotropically sustained sphere of matter.\\

As it should be, the Weyl quadratic invariant identically vanishes in the
Schwarzschild homogeneous model ($\rho=constant$, $p_t=p_r$) but varies indeed like the square
of the tangential pressure in the case of Lemaître's vaults ($p_r=0$). Following our conjecture, we expect therefore a singularity at the place where the
tangential pressure ($p_t>p_r$) diverges.\\
For illustration, let us consider the simple ansatz of Bowers and
Liang \cite{equilibrium}:
\begin{equation}
\label{met2}
e^{2\nu(r)}=\left[\frac{3}{2}\left(1-\frac{r_S}{r_1}\right)^q -\frac{1}{2}\left(1-\frac{r^2}{\mathcal{R}^2}\right)^q\right]^{\frac{1}{q}}
\end{equation}
and
\begin{equation}
\label{boli}
p_r(r)=\rho_0\displaystyle{\frac{\left(1-\frac{r^2}{\mathcal{R}^2}\right)^{q}-\left(1-\frac{r_S}{r_1}\right)^{q}}{3\left(1-\frac{r_S}{r_1}\right)^{q}
-\left(1-\frac{r^2}{\mathcal{R}^2}\right)^{q}}}
\end{equation}
consistent with Eqs.(\ref{lambda}) and (\ref{nup}). The third differential equation given in Eq.(\ref{tov}) implies then 
\begin{equation}
\label{state}
p_t-p_r=\frac{1}{2\rho_0}(\frac{1}{2}-q)(3 p_r+\rho_0)(p_r+\rho_0)
\frac{r^2}{\mathcal{R}^2}\left(1-\frac{r^2}{\mathcal{R}^{2}}\right)^{-1},
\end{equation}
such that the Schwarzschild (Eqs.(\ref{nusch},\ref{psch})) and Lemaître 
(Eqs.(\ref{enu1},\ref{pt1})) models are easily recovered in the limits $q\rightarrow\frac{1}{2}$ and $q\rightarrow 0$, respectively. For the intermediate
case $p_t\ge p_r$ under scruting, a more careful study of Eqs.(\ref{boli}) and
(\ref{state}) reveals (see, for instance, Fig.1) that instability always occurs \underline{first} at the center of the sphere if $0<q\le 1/2\cdot$
Indeed, for $q\ne 0$, the tangential pressure diverges at $r=0$ when the radius $r_1$ tends
to the following critical value
\begin{equation}
\label{r1min}
r_{1}^{MIN}=\frac{r_S}{1-\left(\frac{1}{3}\right)^{\frac{1}{q}}}\cdot
\end{equation}
This interpolation of Eqs.(13) and (17) is nicely confirmed by a numerical
analysis of the Weyl quadratic invariant based on Eqs.(23) and (26) and presented in Fig.\ref{fig2}. Notice that our use of this invariant allows us to
conclude that the glimpse of a naked singularity can only be caught in the Lemaître case ($q=0$). For the other cases ($0<q\le 1/2$), the collapse always begins
at the center where the tangential \underline{and} the radial pressures start to diverge
simultaneously. In fact, tidal forces cancel out at the center $r=0$ 
(see Fig.\ref{fig2}) as expected from the Gauss theorem for a spherically symmetric gravitational field. \\
\section{Gravitational collapse of an incompressible fluid}
In this section we present a simplistic scenario of gravitational
collapse to illustrate the important role played by the
Weyl quadratic invariant in the physics of locally anisotropic spheres.\\

The rather smooth behaviour of the Weyl quadratic invariant (see 
Fig.\ref{fig2} and Eqs.(\ref{weyl2},\ref{state})) suggests that the gravitational collapse of a Schwarzschild supradense body (with $q=1/2$) towards the Lemaître fleeting singularity
($q=0$) might in principle occur under special circumstances. \\
Given the difficulty in deriving realistic equations of
state for anisotropic matter, the common procedure is
to specify an $ad\; hoc$ relation between the radial and tangential pressures.
Let us therefore consider the
collapse of an incompressible fluid and assume that the radial and tangential pressures start to adjust
among themselves in order to remain both finite. Within such a prescription, we preclude an early
implosion once $r_1$ reaches $9/8\;r_S$ (see Fig.3 for illustration). 
In other words, we impose on the Tolman-Oppenheimer-Volkoff equation (\ref{tov})
to be fulfilled at
any time and, thus, at any new position taken by the boundary of the body.\\
At that point, we would like to emphasize that all those
assumptions are rather speculative in the sense that we do
not propose any particular physical mechanism that would lead to them (in particular, the density of the fluid could hardly remain constant in a real
collapse).
However, this set of simplistic hypotheses will help us
in showing the utility of the Weyl quadratic invariant. 
\\
In 
order to maintain the standard equilibrium condition ($0<p_{r,t}<\infty$) despite the gravitational collapse, the radial pressure $p_r$ has to die away
while the tangential pressure decreases
near the center and increases near the surface, making the sphere looking more and more
like a droplet strongly curved by its surface tension (as the anisotropy parameter $q$ is lowered to zero). This change of internal constitution by local anisotropisation of pressures prevents a premature implosion of the sphere.\\
The tidal forces appear first far from the surface 
and grow up as the quasi-static collapse goes on.
The maximum intensity of those tidal forces 
moves then from the heart of the sphere,
when it is still quite isotropically sustained ($q\approx 1/2$ ; $r_1\approx 9/8\; r_S$ ; $p_r\approx p_t$),
to its boundary, when the sphere becomes a Lemaître vault ($q\approx 0$ ; $r_1\approx r_S$ ; $p_r\rightarrow 0$). 
In this way, 
the shrinking body would be more and more tangentially sustained by (local) tidal
forces until the boundary eventually reaches its Schwarzschild radius
where implosion into a black hole necessarily happens (see Fig.\ref{fig4}).\\
A possible astrophysical support
in favour of such a scenario involving
anisotropicaly sustained spheres would of course be the direct observation of redshift factors
\begin{equation}
\label{zmax2}
z^{MAX}=(3)^{\frac{1}{2q}}-1
\end{equation}
much larger than $2$, although not related to the expansion of the Universe.
\section{Conclusion}
By analysing the behaviour of the rather powerful Weyl quadratic invariant
(see Eq.(\ref{weyl1})), we have illustrated how the gravitational collapse
of a spherical body down to its Schwarzschild radius is in principle possible
through the appearance of inner tidal forces whose intensities are proportional
to the local pressure anisotropy. 
\\
Consequently, the Lemaître's original hypothesis concerning tangentially sustained
bodies could be of interest for further investigations in astrophysics
and, in particular, for the study of internal structure of supradense bodies
from the direct observation of sizeable (non-cosmological) redshifts.
Needless to say that precise state
equations have first to be derived in order to proceed. Indeed, such bodies would rather be 
constituted of degenerate nuclear matter with 
locally anisotropic pressures (see \cite{glen}). 
\\
The careful reader has certainly noticed that the analytical expression
given in Eq.(\ref{weyl1}) for the Weyl quadratic invariant is also valid
in the dynamical case where both the metric and matter fields depend on
time $t$ and radial coordinate $r$.
Further work on gravitational
collapse of locally anisotropic spherical bodies in terms of \textit{poles}
in this invariant would therefore be worth pursuing. 
In particular, the regularity of the Weyl quadratic invariant can be used
to characterize the nonsingular asymptotically flat solutions to the
static spherically symmetric Einstein-Yang-Mills equations with $SU(2)$
gauge group.
\subsubsection*{Acknowledgements}
One of the authors ($A.\; F.$) is supported by a grant from the Belgian \textit{Fonds pour la Formation
à la Recherche dans l'Industrie et l'Agriculture} (F.R.I.A.).
\pagebreak

\pagebreak

\begin{figure}
\label{fig1}
\centerline{%
\begin{tabular}{c@{\hspace{2pc}}c}
\includegraphics[scale=0.3,angle=270]{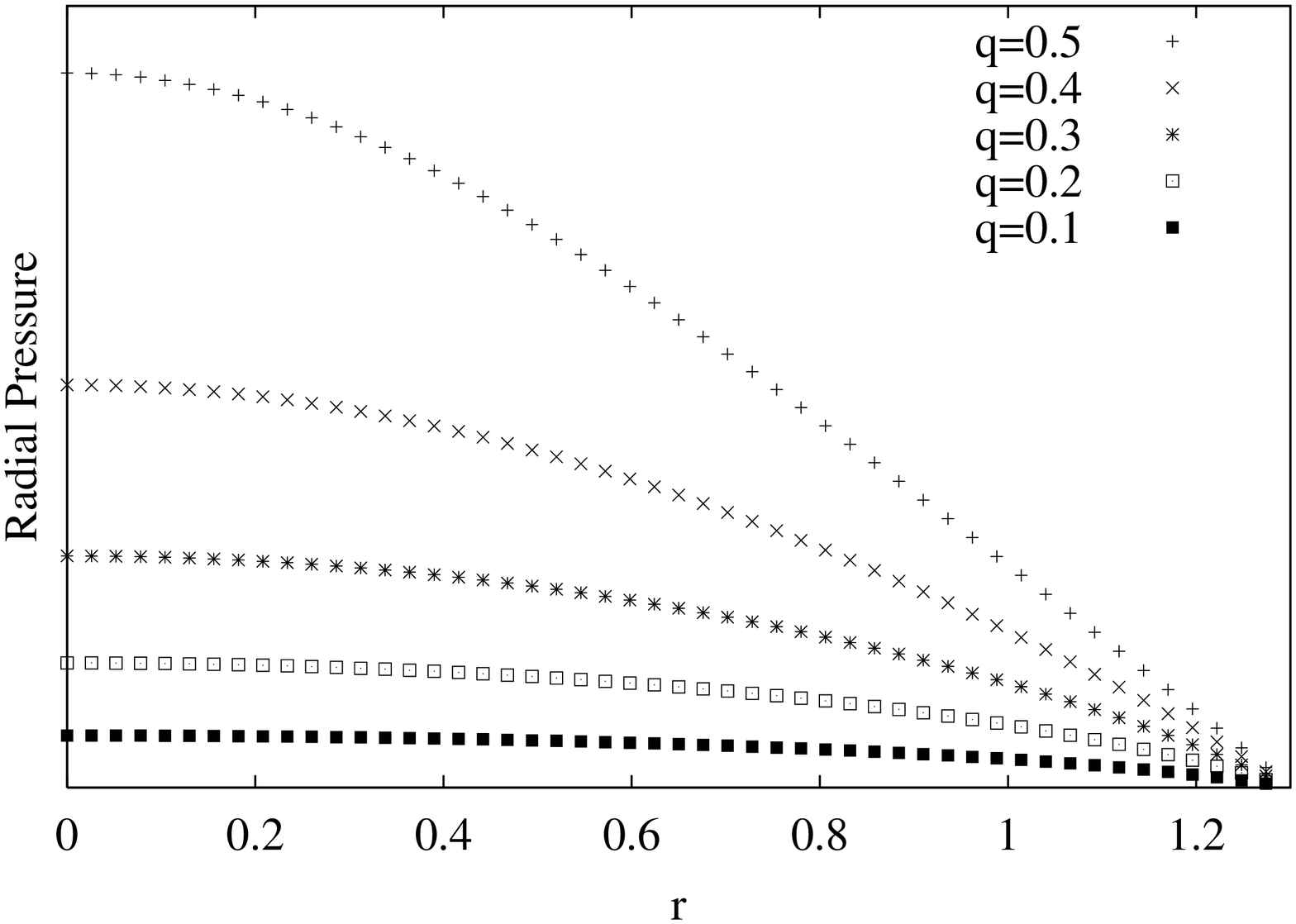} &
\includegraphics[scale=0.3,angle=270]{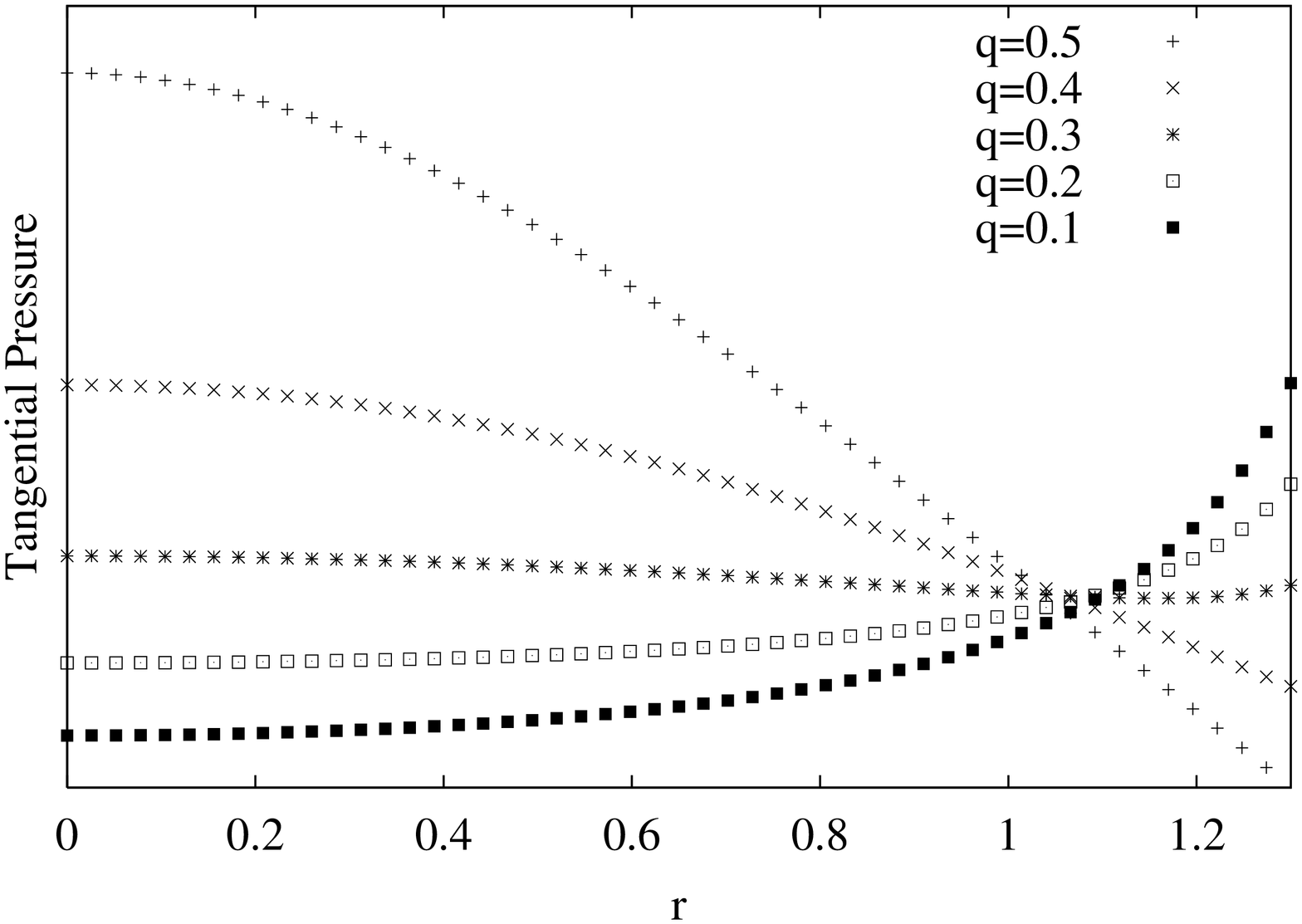}
\end{tabular}}
\caption{\footnotesize \textsl{Radial and Tangential Pressures for different values of the
anisotropy parameter $q$ ($r_1=1.3 r_S$ ; $r_S =1$). } \normalsize}
\end{figure}

\begin{figure}
\begin{center}
\includegraphics[scale=0.3,angle=270]{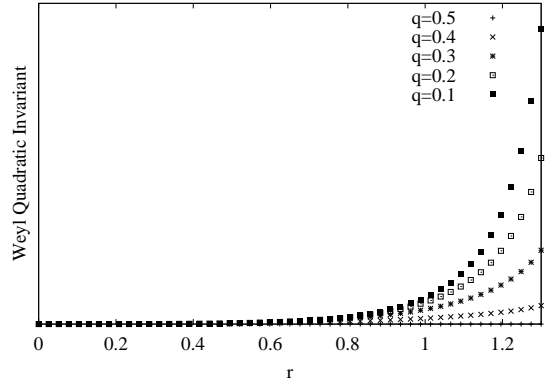}
\caption{\footnotesize \textsl{Weyl Quadratic Invariant $\mathcal{W}$ (see Eq.(\ref{weyl2})), related
to the intensity of tidal effects, for different values of the anisotropy parameter $q$. It is proportional to
the square difference of the two preceding figures ($r_1=1.3 r_S$ ; $r_S =1$).} \normalsize}
\label{fig2}
\end{center}
\end{figure}

\begin{figure}\label{fig3}
\centerline{%
\begin{tabular}{c@{\hspace{2pc}}c}
\includegraphics[scale=0.3,angle=270]{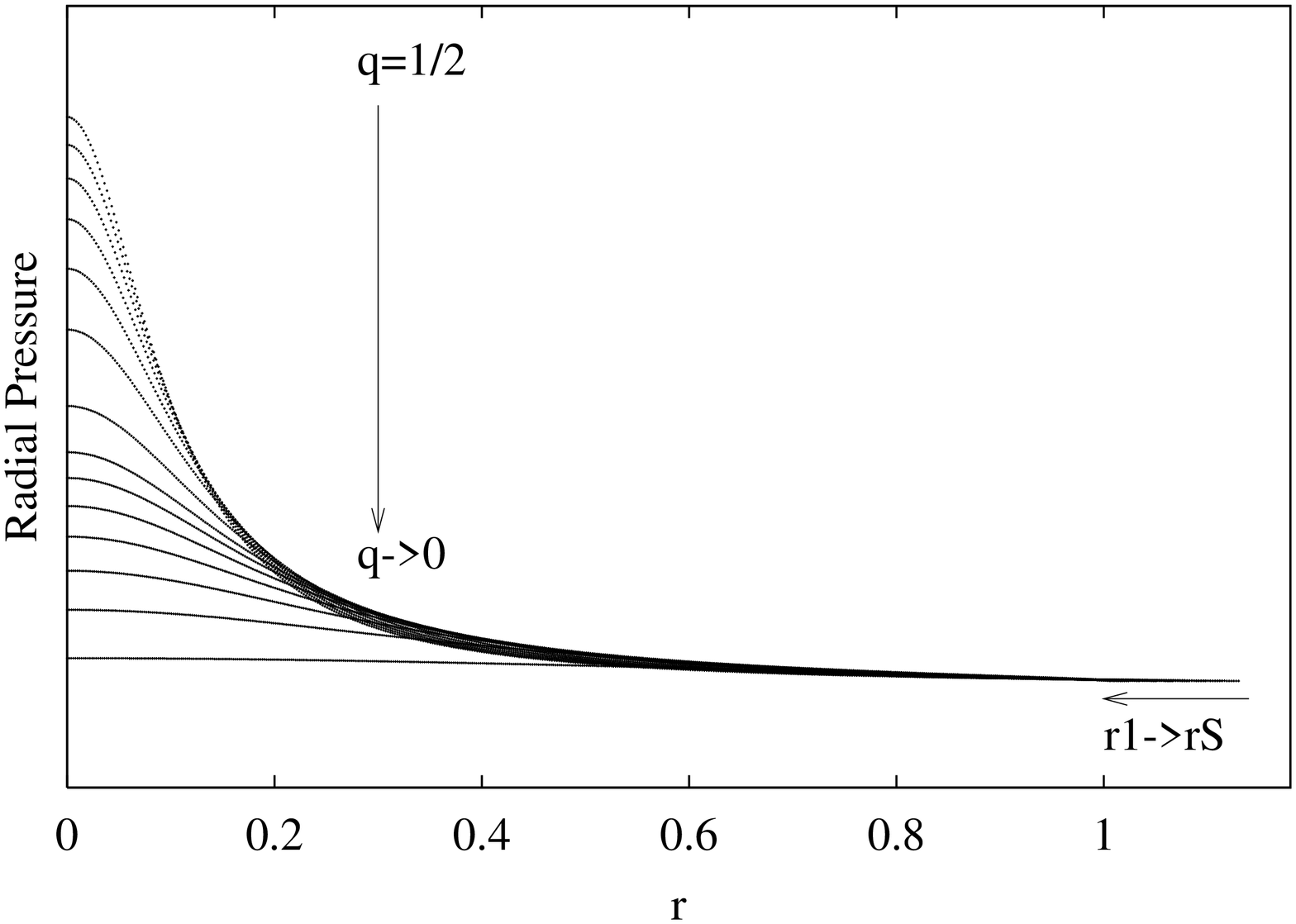} &
\includegraphics[scale=0.3,angle=270]{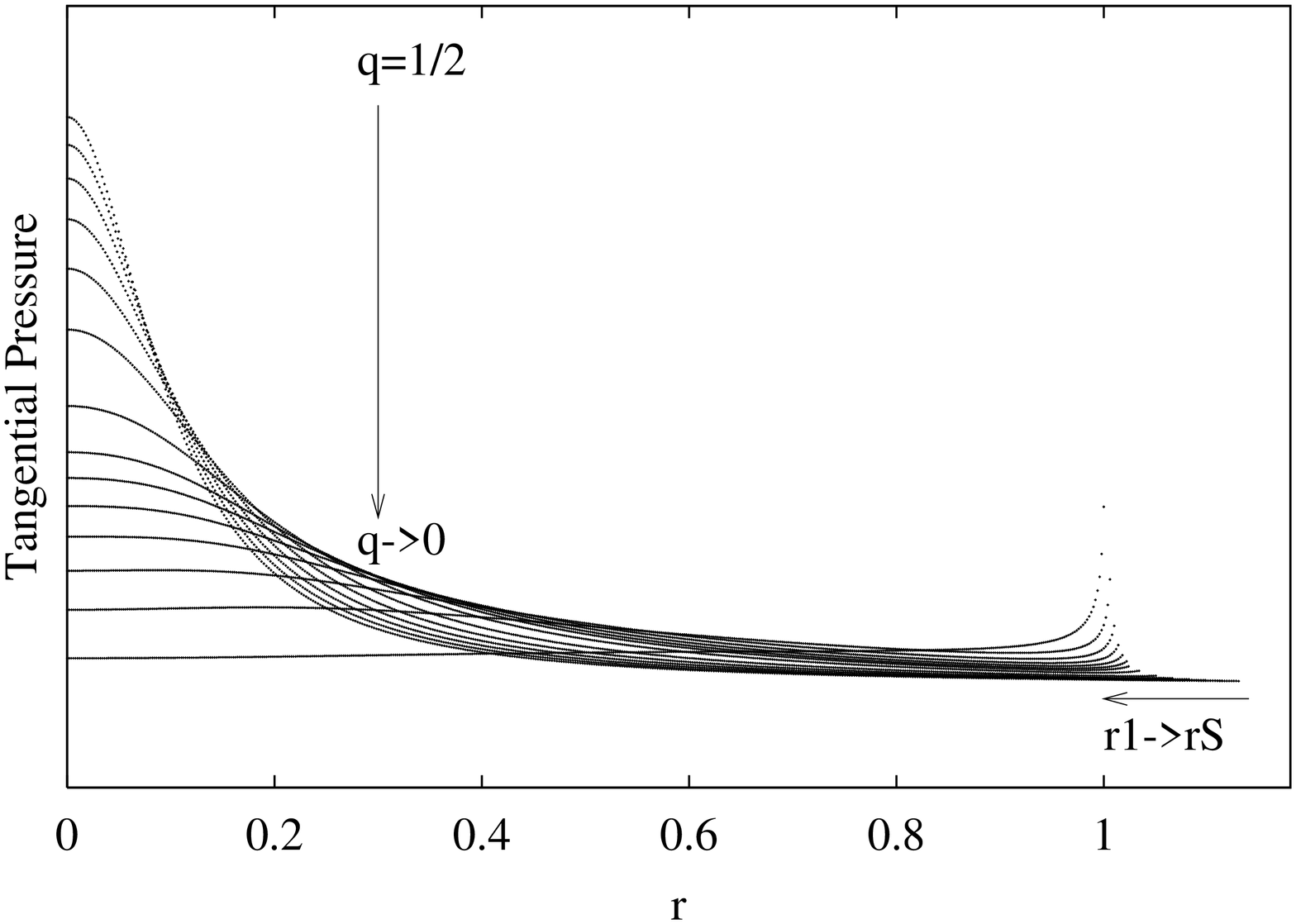}
\end{tabular}}
\caption{\footnotesize \textsl{Modification of the radial and tangential pressures ($p_r$, $p_t$, respectively) needed when
the limit $r_1$ of the sphere goes below the critical radius $\frac{9}{8}r_S$ of the isotropic case where $p_t=p_r$. ($r_S=1$ in the figure.)} \normalsize}
\end{figure}

\begin{figure}
\begin{center}
\includegraphics[scale=0.5,angle=270]{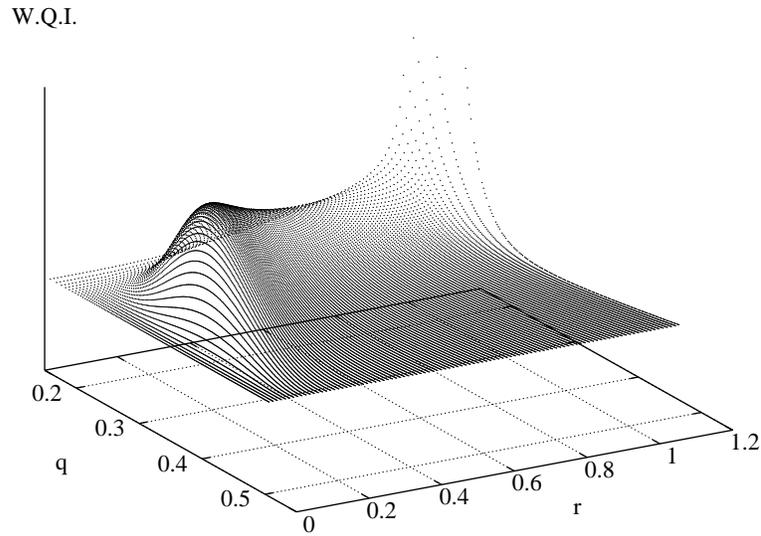}
\caption{\footnotesize \textsl{Weyl Quadratic Invariant ($W.Q.I.$) given by Eq.(\ref{weyl2}) as a function of the anisotropy parameter $q$ and the position $r$ when
the limit $r_1$ of the sphere goes below the critical radius $\frac{9}{8}r_S$ of the isotropic case ($p_t=p_r$). 
($r_S=1$ in the figure.)} \normalsize}
\label{fig4}
\end{center}
\end{figure}

\end{document}